\documentclass[prd,showpacs,preprintnumbers,11pt,,showkeys,onecolumn]{revtex4} % Physical Review D

\usepackage{graphicx}% Include figure files
\usepackage{dcolumn}% Align table columns on decimal point
\usepackage{bm}% bold math
\usepackage{latexsym}
\usepackage{amsfonts}
\usepackage{amssymb}
\usepackage{amsmath}
\usepackage[usenames]{color}
\newcommand{\be}{\begin{equation}}
\newcommand{\ee}{\end{equation}}
\newcommand{\bea}{\begin{eqnarray}}
\newcommand{\eea}{\end{eqnarray}}

\begin{document}
%\preprint{XXX}
%\preprint{YYYY}

\title{ Holographic superconductors with Weyl Corrections via gauge/gravity duality}
\author{D. Momeni$^{1)}$}
\author{R. Myrzakulov$^{1)}$}
\author{M. Raza$^{2)}$}
\affiliation{1)Eurasian International Center for Theoretical Physics,
Eurasian National University, Astana 010008, Kazakhstan\\
2)Department of Mathematics, COMSATS Institute of Information
Technology, Sahiwal, Pakistan}

\date{\normalsize{\emph{Eurasian International Center for Theoretical Physics,}}\\
    \normalsize{\emph{ Eurasian National University, Astana 010008, Kazakhstan}}}

\date{\today}% It is always \today, today,

%===============================================================%
%************************* ABSTRACT ****************************%
%===============================================================%
\begin{abstract}
In this paper, we analytically compute the basic parameters  of the p-wave holographic superconductors with Weyl geometrical corrections using the matching
method. The explicit correspondence between the critical temperature $T_c$ and the dual
charge density $\rho$ has been calculated as
$T_c\propto\rho^{\frac{1}{3}}$ and the dependence of the vacuum expectation
value for the dual condensate operator $\cal{O}$ on the temperature has been found
analytically in the form  $\langle{\cal O}_{+}\rangle\propto
T_c^{\frac{3}{2}}T^{\Delta-\frac{1}{2}}\sqrt{1-(\frac{T}{T_c})^3}$. The
critical exponent 
$\frac{1}{2}$ is an universal quantity according to  predictions of the mean field theory
and independent from the Weyl coupling $\gamma$. Our
analytical results  confirm the  numerical
results and also agree on computations using by the variational method.
\end{abstract}

%Use showkeys class option if keyword
\maketitle

\section{General remarks on gauge-gravity duality}
Maldacena discovered direct relation between a typical coformal field theory
which is invariance under conformal metric transformations on the
boundary, and a gravitational description of a higher dimension
model. This is the simplest interpretation of that so called anti de
Sitter/conformal field theory (AdS/CFT) correspondence conjecture
\cite{maldacena}. As a brief review, we mention the physical basis
for this conjecture which has a vital role in our paper. Firstly, we
discuss conformal field theory (CFT).  By CFT, we mean a quantum
field theory which in it's Hamiltonian form remains conformal
invariant under metric transformations as: \bea
g_{\mu\nu}\rightarrow \Omega(x)g_{\mu\nu}. \eea Now consider a
general transformation of local coordinate $x^{\mu}$ as the
following: \bea x^{\mu}\rightarrow x^{\mu}+u^{\mu}. \eea Here
$u^{\mu}$ is composed of three parts; boost, rotation (Lorentz) and
scaling. The quantum current vector is the solution of the vector
equation $T^{\mu\nu}u_{\nu}=J^{\mu}$, where $T^{\mu\nu}$ is CFT
energy momentum tensor. The conformal field theory energy momentum
tensor has zero trace, it means $T^{\mu}_{\mu}=0$ due to the conservation law
$\partial_{\mu}J^{\mu}=0$. Moreover, it satisfies a local
conservation law. One of the most popular usage of AdS/CFT is a way
to compute the central charge which is an efficient tool for
counting degree of freedoms of the CFT. Three kinds of the
generators have beeb defined associated with conformal
transformations: boost, rotation and scaling or dilatation ones. The
last one, dilatation operator denoted by $D$ is generator of scale
transformations. It acts on any arbitrary unitary operator of CFT in
the form of $D\mathcal{O}=\Delta\mathcal{O}$. Scale invariant
quantum theories are the end point of the renormalization group
flows. Under scaling transformations the CFT operator $\mathcal{O}$
transforms to the new one
$\hat{\mathcal{O}}=\lambda^D\mathcal{O}(\lambda x)$. \par Operators
in CFT are divided into three classes: relevant to the operator
appears in the CFT Hamiltonian $H_{CFT}$ and irrelevant and marginal
ones that they don't appear in the $H_{CFT}$. The key point of the
CFT is that we use the relevant operators on the boundary
$\hat{\mathcal{O}}$ to form the renormalizable  $H_{CFT}$. In this
case we can write the action of a typical field theory as the
following:
\begin{eqnarray}
S=S_{free}+\int d^{d+1}x\Sigma_{i}g_i\hat{\mathcal{O}}_i.
\end{eqnarray}
Here, the first term is the action formed by the free fields and the
integral part contains all interactions and $g_i$ denotes the
coupling constants. By renormalization scheme, we mean the coupling
constants $g_i=g_i(\epsilon)$ satisfies the following integrable
system of differential equations: \bea
\frac{dg_i}{d\log\epsilon}=\beta_i(g_j). \eea The above equation
defines the family of renormalization group (RG) flow equations.
\par Fixed points  of the system are the critical points in which
$\beta_i(g^*_j)=0$. The main property of these fixed points is at
these points the quantum theories become scale invariant and the
unique quantum description of the system is a quantum field theory
in which it remains invariance under conformal transformations, i.e.
CFT. If we perturb CFT we induce another RG flow. In other language,
AdS/CFT relates a gauge theory  on the boundary with dimension $d$,
described by a CFT to gravity theory of matter fields plus black
holes in $d+1$ dimension. So, by gauge gravity we mean \bea
\emph{Gravity}\sim\emph{CFT} \eea The above equivalence is valid in
the t'Hooft large coupling limit in which we assume that the
coupling of the Yang-Mills field $g_{YM}^2N\rightarrow\infty$. We
use gauge-gravity duality via AdS/CFT when the field theories be
very strongly coupled. For finite coupling corrections in the bulk,
we need a full string theory.\par In the language of the statistical physics of thermodynamic systems in thermal equilibrium,there is a simple but very deep relation between partition function
of CFT and the total action of gravity in bulk. It summarizes as the
following relation: \bea Z_{CFT}=e^{-S_{g}} \eea Here $S_g$ is the
gravitational (Euclidean) action (AdS). Moreover, $Z_{CFT}$ denotes
the partition function of the quantum fields calculated by CFT. The
gravity action $S_g$ must be renormalized in the bulk to avoid the
divergences. By using the statistical approach ,it is adequate to
identify the temperature of the gravity (bulk), which is black hole
temperature as the temperature of CFT, so: \bea T_{BH}=T_{CFT }\eea
By definition a holographic superconductor is a superconducting
system under second order phase transition which has a gravity dual
model, via AdS/CFT. The most simplest case is a model of scalar
field $\phi$ with the following action: \bea S\sim\int
d^4x\sqrt{-g}\Big(\mathcal{L}_{CFT}+\mathcal{L}_m\Big) \eea Here
$\mathcal{L}_{CFT}=R+6$ in units of $l_{AdS}=1$. When we add extra
fields with $\mathcal{L}_m$ they induce relevant operators in the
field theory at the asymptotic AdS boundary. This deformation of the field theory
induce an RG flow. Now the full Hamiltonian of the CFT system by
counting the relevant operators represented by: \bea
\tilde{H}=H_{CFT}+\lambda_r\mathcal{O}_r+\lambda_i\mathcal{O}_i \eea
Here operators identifies the fields through their expectation
values, in the asymptotic forms as the following: \bea
\phi=\lambda_r r^{d-\Delta_{\mathcal{O}_r}}+<\mathcal{O}_r>r^{-\Delta_{\mathcal{O}_r}}+..\\
\psi=\lambda_i
r^{d-\Delta_{\mathcal{O}_i}}+<\mathcal{O}_i>r^{-\Delta_{\mathcal{O}_i}}+..
\eea The key point is that here for relevant operators always
$<\mathcal{O}_r>\sim\sqrt{T_c-T}$ which is the same as the
prediction of effective models of superconductors. So, by gauge
gravity we able to compute the correspondence quantities.\par In
condensed matter different aspects of this duality investigated
\cite{condencesd1,condencesd2}. If the gravity part is Einstein
gravity, we have a superconducting phase  \cite{super1,super2}. To
study holographic superconductors we must study black holes in an
asymptotically AdS spacetimes and their gravitational duals. In low
temperature limit, an instability happens and it results in a
breaking of the symmetry and the final state is a holographic
superconductor using dual description.\par
 By curvature corrections, we have different higher order corrected superconductors \cite{GR1,GR2,GB1,GB2,GB3,GB4,GB5,HL1,HL2,wen1,wen2,epl1,born1,born2,cs,maxwell,maxwell2,cai2012}. Also, if the relevant operators be a vector, we have p-wave holographic superconductors \cite{pwave1,pwave2,pwave3}. Very recently we prposed a consistence model for holographic superconductors in F(R) gravity as the gravity theory in bulk \cite{F(R)}. Recently, some papers have discussed the existence of the D-wave
holographic superconductors \cite{d-wave1}-\cite{d-wave5}.\par To
solve the field equations analytically, the first pioneering work
was done by Hertzog \cite{herzog}. Analytical methods are perturbation\cite{herzog,kanno}, the Sturm-Liouville  variational method \cite{analytic1,analytic2,ijtp},the Matching method\cite{GB2} .\\
In the present work, we focus only on Weyl corrections to p-wave
holographic superconductors with the Weyl's coupling 
 value $-\frac{1}{16}<\gamma<\frac{1}{24}$ (see Refs.\cite{mpla}-\cite{weyl}).

Our paper is written as follows: In section 2, we present a toy model for the $3+1$ dimensional holographic superconductor. Section 3, is devoted to the existence of superconducting phase. In section 4, we present analytical properties of this type of superconductors. Conclusions and discussions follow
in section 5.

%%%%%%%%%%%%%%%%%%%%%%%%%%%%%%%%%%%%%%%%%%%%%%%%%%%%%%%%%%%%%%%%%%%%%%%%%%%
\section{p-wave superconductors with Weyl corrections}
%%%%%%%%%%%%%%%%%%%%%%%%%%%%%%%%%%%%%%%%%%%%%%%%%%%%%%%%%%%%%%%%%%%%%%%%%%%%%%
Before study the effects of Weyl coupling on p-wave holographic
superconductors we want to clarify more the topic of holographic
superconductors via AdS/CFT. As we mentioned before the Maldacena
conjecture relates the solutions of type IIB superstring theory in
$AdS_5\times S^5$ to the solutions of a super Yang-Mills theory in a
four dimensional spacetime. The equivalence is valid only as a limiting case
of \emph{large t'Hooft coupling }. Further in the limit of large
numbers of colors and these two quantities are related according to
the fundamental equation $\lambda=g_{YM}^2 N$. In fact, the extra
part of the manifold $AdS_5\times S^5$ must be compact. It means a
four (five) dimensional action is indeed the result on integration
over a ten or eleven dimensional spacetime $AdS_5\times S^5$. If we
use a noncompact of geometry instead of $S^5$ then the classical
geometry has AdS as boundary. The role of filed theory (CFT) is how
to read the quantities (expectation values of the dual operators on
boundary) through an isomorphic map between these operatoes and the
bulk fields, namely $\Psi$. The direct and important explicit result
is the vacuum expectation value of such dual quantum operators which
is equal to the two point functions of the bulk scalar field on
$AdS_5$. \par Just for more better illustration, remembering that
the isomorphism is one to one map between any asymptotic value of
the bulk field and a relevant operator in a gauge invariant field
theory in boundary $<\mathcal{O}_i>$. It means also for any
expression of energy momentum tensor of a typical CFT, there is only
one unique geometrical structure of metric $g_{\mu\nu}$ in bulk.
This is a gauge/gravity duality conjecture. Due to the 
 dynamical approach, we have an equivalence relation between \emph{on-shell}
value of the total action of fields in the bulk, super string theory of gravity and the
action of CFT. Here on-shell means if we compute the string action
with the classical solutions of field equations, such equivalence
appears. Just to recall the main idea, we consider here again the
type IIB string (super) theory and it's quantum field theoretical
description via four dimensional CFT. Because, we will use the
concepts of asymptotic value of scalar field as the vacuum
expectation of the CFT operators, here is very usefull to re
calculate explicitly relation between bulk scalar fields two point
functions and expectation value of CFT operators. The computation
starts by solving the scalar field equation of motion for a point
like source in the $AdS_5$ spacetime: \bea
(\nabla_{\mu}\nabla^{\mu}-m^2)\psi=0. \eea Here, mass $m^2$ as we
will observe, is a code for the asymptotic behavior of the field and
also $<\mathcal{O}_i>$. In a simple $AdS_5$ background, the
propagator reads as the following, as we can find in many
references: \bea
G(r;x,x')=\frac{\Gamma(\Delta)}{\pi^2\Gamma(\Delta-2)}(\frac{L^2}{r})^{\Delta}((\frac{L^2}{r})^{2}+|x-x'|^2)^{-\Delta}.
\eea As usual $x'$ denotes retarded coordinate. The key object here
is the following relation between $\Delta$ and $m^2$: \bea
\Delta(\Delta-4)=m^2L^2. \eea The above equation as many people
wrote has two positive roots but usually people take the case of
$\Delta>2$. In fact from unitarity $1<\Delta<4$ and in this case,
the scalar mass is bounded in the BF region.\par Now if we expand
Green function near the AdS boundary in which $r\rightarrow\infty$
we have: \bea
G(r;x,x')\sim\alpha(x)(rL^{-2})^{\Delta-4}+\beta(x)(rL^{-2})^{-\Delta}.
\eea Here as we can check directly, the first function $\alpha$ is
the source of the scalar field $\psi$ evaluated on AdS boundary and
the second $\beta(x)$ is the vacuum expectation value computed for the dual
operator $\mathcal{O}_i$. We are free to choice how we quantize the
operator and it means how we fix the constant factors. In the
holographic superconductors if we fix the form of the operator
(s-wave,p-wave,..) depending on the form of the condensator, then
our main goal is "`how we compute $\beta(x)$ and also $\alpha(x)$"'.
By first we will have the expression of the condensation
$<\mathcal{O}>$. If we put this solutions as on-shell in the bulk
action and doing simple algebra and because we know the action of
string and CFT are the same so we can conclude that two point
function is exactly the same as the kernel of scalar field in bulk.
Here it is needed to identify the color number in terms of the
Newton;s gravitational constant and AdS radius.\par After this brief
mini review of applied AdS/CFT technique we proceed our plan to
study p-wave holographic superconductors . We take the action of a
p-wave superconductor in the following form \cite{epl2}:
 \begin{eqnarray}
S=\int dt d^{4}x\sqrt{-g}\{\frac{1}{16\pi
G_{5}}(R+12)-\frac{1}{4g^2}F^{a}_{\mu\nu}F^{a\mu\nu}+\gamma
C^{\mu\nu\rho\sigma}F^{a}_{\mu\nu}F^{a}_{\rho\sigma}\}.
\end{eqnarray}
The bulk action here is five dimensional integral. We have also
$G_5$ as gravitational coupling . We set the AdS radius $L=1$. The
 field strength components $F^{a}_{\mu\nu}$ are computed using
$A_{\mu}^{a}$'s (non-Abelian gauge fields):
 \begin{eqnarray}
 F^{a}_{\mu\nu}=2\partial_{[\mu}A^{a}_{\nu]}
 +\varepsilon^{abc}A^{b}_{\mu}A^{c}_{\nu},\{a=1,2,3,\mu=0,1,2,3\}.
 \end{eqnarray}
 We introduce the non zero Weyl's coupling $\gamma$ which is a constant and it
is limited inside the interval
$-\frac{1}{16}<\gamma<\frac{1}{24}$.\par To fix the unique geometry
on bulk and as an attempt to find the dual CFT operator's
expectation values, we take the metric as an asymptotic $AdS_5$
background:
\begin{eqnarray}
 ds^2=r^2(-fdt^2+dx^idx_i)+\frac{dr^2}{r^2f},\ \
 i=1,2,3\label{metric}.
\end{eqnarray}
Here the metric function is:
\begin{eqnarray}
f=1-(\frac{r_{+}}{r})^4.
\end{eqnarray}
The  horizon coincides on $r=r_{+}$. The black hole temperature
which is the same as CFT temperature, due to the simple equivalence
between two different free energies of two configurations, reads as
$T=\frac{r_{+}}{\pi }$. The full system of field equations are\cite{weyl}:
\begin{eqnarray}\label{GYME1}
\nabla_{\mu}\left( F^{a\mu\nu} - 4\gamma C^{\mu\nu\rho\sigma}
F^{a}_{\rho\sigma} \right) =-\epsilon^{a}_{bc}A^{b}_{\mu}F^{c\mu\nu}
+ 4\gamma
C^{\mu\nu\rho\sigma}\epsilon^{a}_{bc}A^{b}_{\mu}F^{c}_{\rho\sigma}.
\end{eqnarray}
Now to explain the superconductivity in the p-wave discipline we
take the Yang-Mills gauge field in the form of a two component one
form which is homogeneity but anisotropic: \cite{gauge}:
 \begin{eqnarray}
 A=\phi(r)\sigma^{3}dt+\psi(r)\sigma^1 dx.
 \end{eqnarray}
As usual $\sigma^{i}$ denotes Pauli's matrixes.
 The resulting field  equations Yang-Mills (\ref{metric}) are as the following:
\begin{eqnarray}
\label{EOMr1} \left( 1 - \frac{24 \gamma r_{+}^{4}}{r^{4}}
\right)\phi'' + \left( \frac{3}{r} + \frac{24 \gamma
r_{+}^{4}}{r^{5}} \right)\phi'
- \left( 1 + \frac{8 \gamma r_{+}^{4}}{r^{4}} \right)\frac{\psi^{2}\phi}{r^{4}r_{+}^2f}=0,\\
\label{EOMr2} \left( 1 - \frac{8 \gamma r_{+}^{4}}{r^{4}}
\right)\psi'' + \left[ \frac{3}{r} + \frac{f'}{f} - \frac{8 \gamma
r_{+}^{4}}{r^{4}} \left( -\frac{1}{r} + \frac{f'}{f} \right) \right]
\psi' + \left( 1 + \frac{8 \gamma r_{+}^{4}}{r^{4}} \right)
\frac{\phi^{2}\psi}{r^{4}r_{+}^2f^{2}}=0, \ \ f'=\frac{df}{dr}.
\end{eqnarray}
 To recover
the asymptotic behavior and also to solve the possible problem of
holographic renormalizability, it is adequate to rewriting
(\ref{EOMr1},\ref{EOMr2})  with respect to the new coordinate
$z=\frac{r_{+}}{r}$, it corresponds to the  blackhole horizon $z=1$,
and the  AdS boundary
 at $z=0$. New set of the equations of motion
are:
\begin{eqnarray}
\label{EOMz1} \left( 1 - 24 \gamma z^{4} \right)\phi'' -
\frac{1}{z} \left( 1 + 72 \gamma z^{4} \right)\phi'
- \left( 1 + 8 \gamma z^{4} \right)\frac{\psi^{2}\phi}{r_{+}^2f}=0,\\
\label{EOMz2} \left( 1 - 8 \gamma z^{4} \right)\psi'' + \left[
-\frac{1}{z} + \frac{f'}{f} - 8 \gamma z^{4} \left( \frac{3}{z} +
\frac{f'}{f} \right) \right] \psi' + \left( 1 + 8 \gamma z^{4}
\right) \frac{\phi^{2}\psi}{r_{+}^2f^{2}}=0,\ \ f'=\frac{df}{dz}.
\end{eqnarray}
We present the full numerical results for $T_c$ the critical temperature, in which for $T>T_c$ , when the system evolves dynamically in the normal phase  with $\psi=0$ and for $T<T_c$ we have a YM hairy black hole for $\psi\neq0$ for $m^2=-2$ in the table for different Weyl coupling constants. We used shooting algorithm as well as imposing boundary conditions on the horizon.
\begin{widetext}
\begin{table}[ht]
\begin{center}

\begin{tabular}{|c|c|c|c|c|c|c|}
         \hline
$~~\gamma~~$
&~~$-0.06$~~&~~$-0.04$~~&~~$-0.02$~~&~~$0$~~&~~$0.02$~~&~~$0.04$~~
          \\
       \hline
~~$T_{c}$~~ & ~~$0.1701\rho^{1/3}$~~ & ~~$0.1774\rho^{1/3}$~~ &
~~$0.1869\rho^{1/3}$~~& ~~$0.2005\rho^{1/3}$~~&
~~$0.2239\rho^{1/3}$~~& ~~$0.3185\rho^{1/3}$~~
          \\
        \hline
\end{tabular}
\caption{\label{Tc} The critical temperature $T_{c}$ versus the different
values of Weyl coupling parameter $\gamma$ using Numerical solutions to the field equations \cite{epl2}.}
\end{center}
\end{table}
\end{widetext}
 Our aim here is to recover these analytic results
using the analytical approaches.

\begin{figure*}[thbp]
\includegraphics[width=7.5cm]{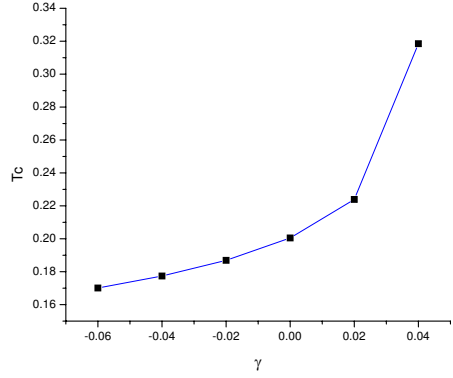}
\caption{ Graph shows the values of $T_c$ as a function of
Weyl's coupling . Here $\rho=1$.}
\end{figure*}
FIG.1 shows that by increasing Weyl coupling $\gamma$, the critical temperature increases so condensation becomes more hard. Increasing scheme is a monotonic function of coupling $\gamma$.

%%%%%%%%%%%%%%%%%%%%%%%%%%%%%%%%%%%%%%%%%%%%%%%%%
\section{Existence of a Superconductor phase}
%%%%%%%%%%%%%%%%%%%%%%%%%%%%%%%%%%%%%%%%%%%%%%%%%%
Before we explained how the superconducting phase happens when the
system undergoes a lower temperature than $T_c$.  In this section we
try to explain superconducting phase using matching
technique. We like to mention here that we have followed largely the
exposition given in \cite{GB1}.\par We start by studying the system
of equations of motion given in (\ref{EOMr1}) and (\ref{EOMr2}). The
first observation is that the system  (\ref{EOMr1}) and
(\ref{EOMr2}) has the following non trivial solution in normal
phase, i.e. when $T<T_c$ and $\psi=0$:
\begin{eqnarray}
\phi = \phi_0(r) &=& \frac{\rho}{r_+^2} \left ( 1 -
\frac{r_+^2}{r^2} \right ),\ \ \rho=\mu r_{+}^2, \\ \psi & \equiv &
0.
\end{eqnarray}
One can verify this solution by examining the field equations in
normal phase of $\psi=0$.  The physical meaning of these solutions
is the existence of the normal phase in which no condensation
happens. We will now prove that there is a specific critical
temperature $T_c$ that system under it goes to superconducting
phase. It is useful to rewrite the (\ref{EOMr1}) as:
\begin{eqnarray}
\frac{d}{dr}\Big[(-r^3+24\gamma\frac{r_{+}^4}{r})\phi'\Big]+(1+8\gamma\frac{r_{+}^4}{r^4})\frac{\psi^{2}\phi}{rr_{+}^2f}=0\label{pertubphi}.
\end{eqnarray}
In the normal phase $T>T_c$  when $\psi=0$, the zero-order solution
satisfies
 \bea
\frac{d}{dr}\Big[(-r^3+24\gamma\frac{r_{+}^4}{r})\phi_0'\Big]=0.
 \eea
As a usual method for study instabilities of the models, we perturb
$\phi(r) = \phi_0(r) + \delta \phi$. Then (\ref{pertubphi}) up to
the first-order perturbations shows: \be
\frac{d}{dr}\Big[(-r^3+24\gamma\frac{r_{+}^4}{r})\delta\phi'\Big]
\geq 0, \ee In the AdS boundary when the fields tend to their values
on boundary surface, \bea (-r^3+24\gamma\frac{r_{+}^4}{r}) \phi'\to
-r^3\phi' \to -2\rho = -r^3\phi_0' ,\eea
 consequently,
\bea (-r^3+24\gamma\frac{r_{+}^4}{r}) \delta\phi' \to 0. \eea At AdS
boundary,with aim of  $\delta\phi |_{r_+}= 0\Longrightarrow \delta
\phi' \leq 0 $. So \be \phi(r) \leq \phi_0(r). \
 \ee

Now we consider second component of the gauge field, and we write it in the self-adjoint form:
 \be
\frac{d}{dr}\Big[(-r^3+24\gamma\frac{r_{+}^4}{r})\psi'\Big]-(1+8\gamma\frac{r_{+}^4}{r^4})\frac{\psi^{2}\phi}{rr_{+}^2f^2}=0\label{pertubpsi}.
\ee
 We define a new variable $X=r\psi$. Now, the boundary condition
$\psi'(r_+)=0$ at the horizon implies that the new field satisfies $X'_+ = X_+/r_+$. Further in the asymptotic limit, $rfX' \to 0$, so the existence of a condensatation state needs
 a typical critical point in field function$X$, 
\be
X'(r_{_T})=0\wedge X''<0\wedge X>0.
\ee
Using $\phi_0(r)$ we have
qualitatively verified that p-wave Weyl superconductors exist. 
%%%%%%%%%%%%%%%%%%%%%%%%%%%%%%%%%%%%%%%%%%%%%%%%%%%%%%%%%%%%%%%%%%%%%%%%%%%%
\section{Critical temperature and condensation values by matching method}
%%%%%%%%%%%%%%%%%%%%%%%%%%%%%%%%%%%%%%%%%%%%%%%%%%%%%%%%%%%%%%%%%%%%%%%%%%%%%%
 Because we
showed that before, if we start from a scalar field model in the
bulk, the on-shell stringy action, evaluated by the solutions of
this scalar field has a kernel(Green function) which is exactly the
same as the two point function of a relevant operator namely as
${\cal O}$ in the dual CFT.\par Thus, we are ready to compute
analytically the condensate $\langle {\cal O} \rangle$ with fixed
$\rho$. We start by studying the boundary conditions in $z$
coordinate. Regularity at  $z=1$,
requires
\begin{eqnarray}
\phi(1)=0\,,\hspace{1cm}\psi^\prime(1)=0\,,
\label{regularity}
\end{eqnarray}

The matching method starts by writing the series solution of the
fields around the boundary points $z=1$ and $z=0$. We assume that
these fields are non singular. So, the usual power series work for
us. The next step is to match these two families smoothly. It means
we suppose that here, fields and their first derivatives remain
continuous. We have no logical reasons to have a jump in these
boundary quantities.

%%%%%%%%%%%%%%%%%%%%%%%%%%%%%%%%%%%%%%%%%%%
\subsection{Solution near the horizon}
%%%%%%%%%%%%%%%%%%%%%%%%%%%%%%%%%%%%%%%%%%%
We expand gauge fields $\{\phi,\psi\}$ near $z=1$ as
\begin{eqnarray}
\phi(z)&=&\phi(1)-\phi^\prime(1)(1-z)+\frac{1}{2}\phi^{\prime\prime}(1)(1-z)^2
+\cdots\label{series1},\\
\psi(z)&=&\psi(1)-\psi^\prime(1)(1-z)+\frac{1}{2}\psi^{\prime\prime}(1)(1-z)^2
+\cdots\label{series2}.
\end{eqnarray}
We set $a=-\phi^\prime(1)<0$ and
$b=\psi(1)>0$ . For the 2$^{\rm nd}$ order coefficient $\phi$ we have
(\ref{EOMz1}):
\begin{eqnarray}
\phi''(1)&=&-a\frac{1+72\gamma}{1-24\gamma}+\frac{1+8\gamma}{1-24\gamma}
\frac{ab^2}{4r_{+}^2}.
\end{eqnarray}
So (\ref{series1}) reads as
\begin{eqnarray}
\phi(z)=a\Big[(1-z)-\frac{(1-z)^2}{2(1-24\gamma)}\Big(1+72\gamma-\frac{b^2}{4r_{+}^2}(1+8\gamma)\Big)\Big]\label{seriesphi}.
\end{eqnarray}
For $\psi$ similarly , via
(\ref{EOMz2}) as
\begin{eqnarray}
\psi''(1)&=&-\frac{1}{32r_{+}^2}\Big(\frac{1+8\gamma}{1-8\gamma}\Big)a^2b.
\end{eqnarray}
So, we write the following series solution for scalar field
(\ref{series2})
\begin{eqnarray}
\psi(z)&=&b\Big[1-\Big(\frac{1+8\gamma}{1-8\gamma}\Big)\frac{(1-z)^2}{64r_{+}^2}a^2\Big]\label{seriespsi}.
\end{eqnarray}

\subsection{Solution near the asymptotic AdS region}
From (\ref{EOMz1}) and (\ref{EOMz2}), in the asymptotic region we
have
\begin{eqnarray}
\phi(z)&=&\mu-\frac{\rho}{r_{+}^2}z^2\label{ads1},\\
\psi(z)&=&\frac{<\cal O_{+}>}{r_{+}^\Delta}z^\Delta\label{ads2}.
\end{eqnarray}
For renormalizability,we set $<\mathcal O_{-}>=0$. We clarify this quantization scheme as the following. A simple check shows that  under  mass bound, only  $\Delta_{+}$ has an enough rapid fall off and the scalar field behaves as
\begin{eqnarray}
\psi(r)\rightarrow   r^{-(\Delta_{+}+2)}+ r^{-\Delta_{+}}<\cal O_{+}>\,.
\end{eqnarray}
It means that the scalar field is dual to a quantum operator  on the boundary with conformal dimension $\Delta_{+}$ and we can ignore $\cal O_{-}$.
 This is not the unique possible choice. It is easy to find that if
mass be out of the BF bound, both of the terms with conformal
dimensions $\Delta_{\pm}$ fall off and we can keep them.  In
conclusion, the quantization scheme is a valid procedure. In any
case, the scalar field is asymptotic to  $<\cal{O_{\pm}}>$ and these
are dual to  operators with dimension $\Delta_{\pm}$. However, in
this work  we limit ourselves to the fall off with $\Delta_{+}$.

\subsection{Matching and phase transition}
Now, we smoothly join the solutions (\ref{seriesphi}) and
(\ref{seriespsi}) with (\ref{ads1}) and (\ref{ads2}) respectively at
$z=z_m$. Smoothly continuty requires \bea
\phi^{z=0}|_{z_m}=\phi^{z=1}|_{z_m},\ \ \phi'^{z=0}|_{z_m}=\phi'^{z=1}|_{z_m}\\
\psi^{z=0}|_{z_m}=\psi^{z=1}|_{z_m},\ \ \psi'^{z=0}|_{z_m}=\psi'^{z=1}|_{z_m} \eea So,
explicitly we obtain:
\begin{eqnarray}
&&\mu-{\frac {\rho\,{z_{{m}}}^{2}}{{\pi }^{2}{T}^{2}}}=a \left(
1-z_{{m} }-\frac{1}{2}\,{\frac { \left( 1-z_{{m}} \right) ^{2}
\left( 1+72\,\gamma-\frac{1}{4r_{+}^2} \,{b}^{2} \left( 1+8\,\gamma
\right)  \right) }{1-24\,\gamma}}
 \right),
 \label{c:phi}\\
&&-\,{\frac {2\rho\,z_{{m}}}{{\pi }^{2}{T}^{2}}}=a \left( -1+{\frac {
 \left( 1-z_{{m}} \right)  \left( 1+72\,\gamma-\frac{1}{4r_{+}^2}\,{b}^{2} \left( 1+8
\,\gamma \right)  \right) }{1-24\,\gamma}} \right),
 \label{c:dphi}\\
&&\frac {{z_{{m}}}^{\Delta}<\cal O_{+}>}{(\pi T)^{\Delta}}=b \left(
1-{ \frac {1}{64r_{+}^2}}\,{\frac { \left( 1+8\,\gamma \right)
{a}^{2} \left( 1-z _{{m}} \right) ^{2}}{1-8\,\gamma}} \right),
 \label{c:psi}\\
&&\,{\frac {\Delta z_{m}^{\Delta-1}<\cal O_{+}>}{(\pi T)^{\Delta}}}=\frac{1}{32r_{+}^2}\,{
\frac {b \left( 1+8\,\gamma \right) {a}^{2} \left( 1-z_{{m}} \right) }
{1-8\,\gamma}}
 \label{c:dpsi}.
\end{eqnarray}
Recalling that $r_+=\pi T$, after eliminating $a b^2$ from
(\ref{c:phi}) and (\ref{c:dphi}) gives
\begin{eqnarray}
\mu&=&\frac{1}{2}\,{\frac {2\,\rho\,z_{{m}}-az_{{m}}{\pi
}^{2}{T}^{2}+a{\pi }^ {2}{T}^{2}}{{\pi }^{2}{T}^{2}}},
\\
b&=&\sqrt {-{\frac
{-8\,\rho\,z_{{m}}+192\,\rho\,z_{{m}}\gamma-384\,a \gamma\,{\pi
}^{2}{T}^{2}+4\,az_{{m}}{\pi }^{2}{T}^{2}+288\,az_{{m}} \gamma\,{\pi
}^{2}{T}^{2}}{a(1+8\,\gamma-z_{{m}}-8\,z_{{m}}\gamma)}}} \label{b}.
\end{eqnarray}
 We
eliminate the $a^2b$ term from (\ref{c:psi}) and (\ref{c:dpsi}) and obtain
\begin{eqnarray}
<\cal O_{+}>&=&\frac { \left( \pi \,T \right) ^{\Delta}b}{{z_{{m}}}^{\Delta-1}
 \left( \frac{1}{2}\,\Delta+z_{{m}} \left( 1-\frac{1}{2}\,\Delta \right)  \right) }
  \ . \label{relation:2}
\end{eqnarray}

For non-vanishing b, we can eliminate $<\cal O_{+}>$ from (\ref{c:psi}) , (\ref{c:dpsi}) to obtain
\begin{eqnarray}
a&=&8\pi T{\frac { \,\sqrt { \left(1- z_{{m}} \right)  \left( -\Delta-2\,
z_{{m}}+z_{{m}}\Delta \right)  \left( 1+8\,\gamma \right) \Delta\,
 \left( -1+8\,\gamma \right) }}{ \left( z_{{m}}-1 \right)  \left( -
\Delta-2\,z_{{m}}+z_{{m}}\Delta \right)  \left( 1+8\,\gamma \right)
}}.
\end{eqnarray}

With this $a$, from (\ref{b}) we obtain

\begin{eqnarray}
b&=& \sqrt {36\left(  \left( \gamma+{\frac {1}{72}} \right) z_{{m}}-\frac{4}{3}\,\gamma
 \right) {\pi }^{3}\sqrt { (1- z_{{m}})  \left( -\Delta-2
\,z_{{m}}+z_{{m}}\Delta \right)  \left( 1+8\,\gamma \right) \Delta\,
 \left( -1+8\,\gamma \right) }
}\nonumber\\&& \times\sqrt {{\frac {{T}^{3} -T_c^3 }{ \left( \frac{1}{8}+\gamma \right)  \left( -1+z_{{m}} \right)
T\sqrt {\left( 1-z_{{m}} \right)  \left( 1+8\,\gamma \right)
 \left( -1+8\,\gamma \right) \Delta\, \left( -2\,z_{{m}}+\Delta\,z_{{m
}}-\Delta \right) }\pi }}}.
\end{eqnarray}
$T_c$ is defined  as
\begin{eqnarray}
T_c&=&\alpha(\gamma|z_m|\Delta)\rho^{\frac{1}{3}}\label{Tc}
\end{eqnarray}
With
\begin{eqnarray}
\alpha(\gamma|z_m|\Delta)&=&\frac{1}{2}\,\sqrt
[3]{2(1+8\,\gamma)(-1+24\,\gamma)}\times\sqrt [3]{\frac
{z_m}{z_m(1+72\gamma)-96\gamma}\sqrt{\frac{(1-z_m)(-\Delta+z_m(\Delta-2))}{\Delta}}}
\end{eqnarray}
Here, as we know $-\frac{1}{16}<\gamma<\frac{1}{24}$. For positivity
of the $T_c$, we have two possibilities (remembering that here the
conformal dimension can be negative )

\begin{itemize}
\item Case $A:\ -\frac{1}{16}<\gamma<-\frac{1}{8}, \gamma>\frac{z_m}{96-72z_m}$,
\item Case $B:\ -\frac{1}{8}<\gamma<\frac{1}{24},
\gamma<\frac{z_m}{96-72z_m}$.
\end{itemize}
In both cases, the conformal dimension reads as $\Delta>-\frac{2z_m}{1-z_m}$.

\begin{figure*}[thbp]
\begin{tabular}{rl}
\includegraphics[width=5.5cm]{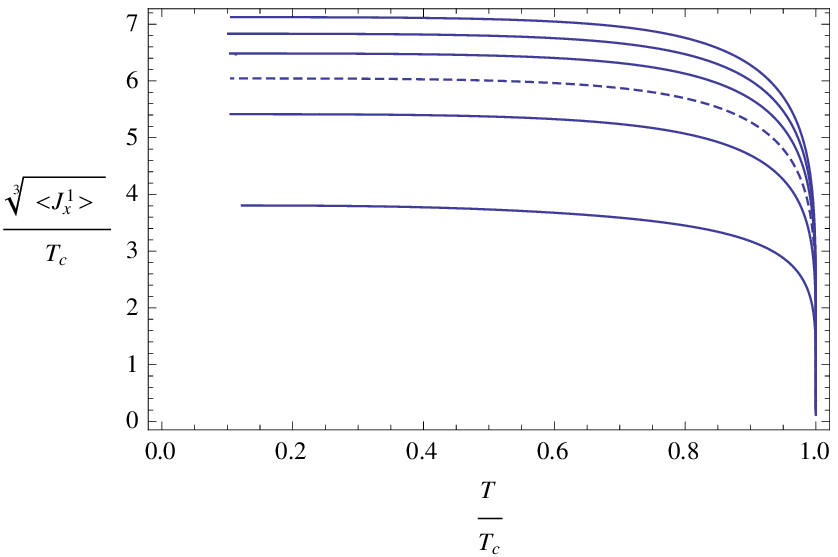}&
\includegraphics[width=4.5cm]{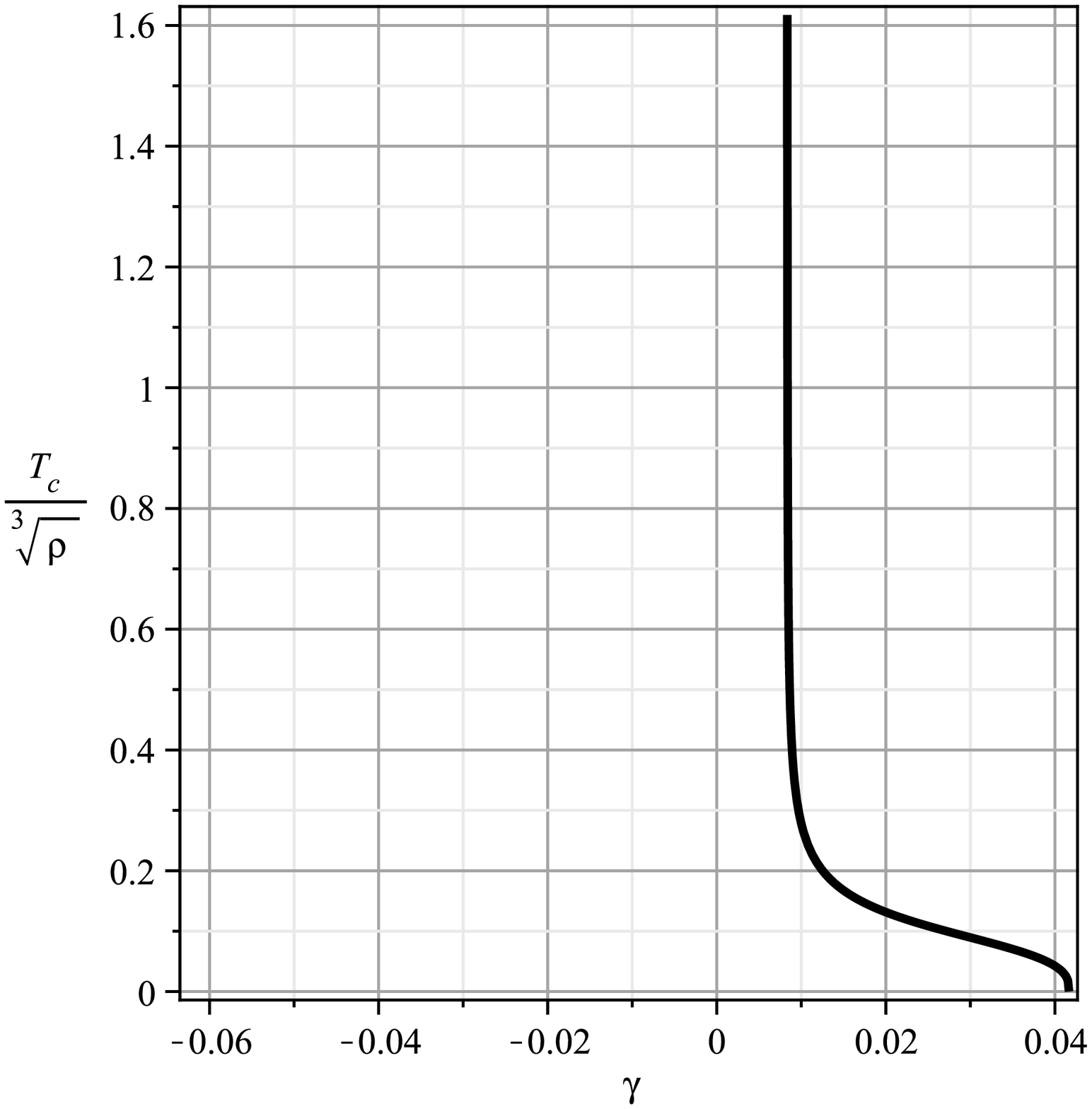}
\includegraphics[width=5.5cm]{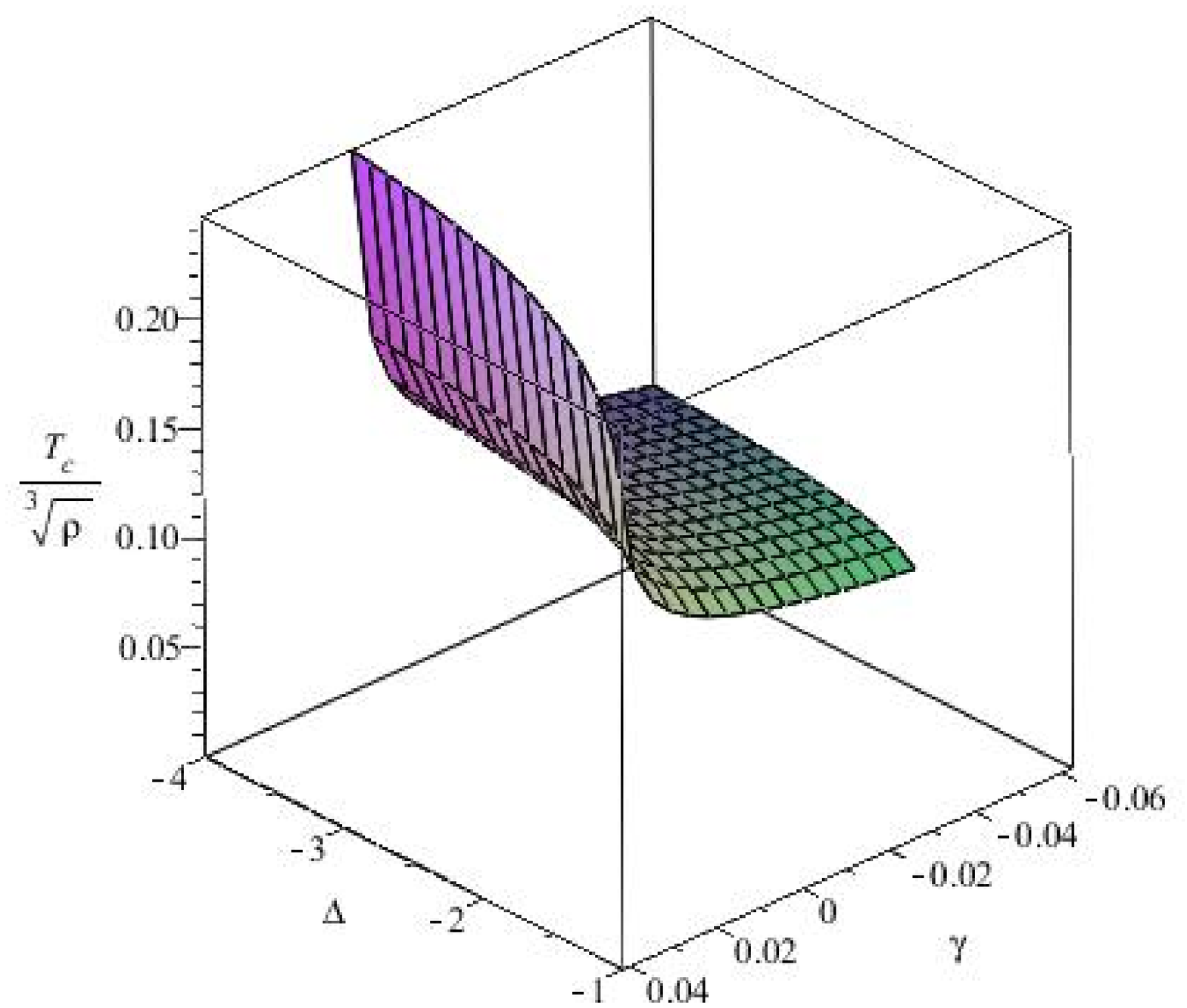} \\
\end{tabular}
\caption{ (\textit{Left})Numerical result shows the condensation $<{\cal O_{+}}>=<J_x^1>$ as a function of
temperature.  (\textit{Middle}) Analytical result shows the  $\frac{T_c}{\rho^{\frac{1}{3}}}$ as a function of
the Weyl's coupling  for $z_m=\frac{1}{2}$,$\Delta=3$..
   (\textit{Right}) 3D plot of  the  normalized $T_c\equiv\frac{T_c}{\rho^{\frac{1}{3}}}$ as a function of
the $\gamma,\Delta$ for $z_m=\frac{1}{2}$.}
\end{figure*}
 Finally 
$\langle{\cal O}_+\rangle$
is:
\begin{eqnarray}
\langle{\cal O}_{+}\rangle
=\beta(\gamma|z_m|\Delta)T_c^{\frac{3}{2}}T^{\Delta-\frac{1}{2}}\sqrt{1-(\frac{T}{T_c})^3},
\end{eqnarray}
where
\begin{eqnarray}
\beta(\gamma|z_m|\Delta)&=&\frac{\sqrt {36\left(  \left( \gamma+{\frac {1}{72}} \right) z_{{m}}-\frac{4}{3}\,\gamma
 \right) {\pi }^{3}\sqrt {( 1-z_{{m}} ) ( -\Delta-2
\,z_{{m}}+z_{{m}}\Delta)( 1+8\,\gamma) \Delta\,
 ( -1+8\,\gamma) }
}}{{z_{{m}}}^{\Delta-1}
( \frac{1}{2}\,\Delta+z_{{m}} ( 1-\frac{1}{2}\,\Delta ))}\nonumber\\&& \times
\frac{\pi^{\Delta}}{\sqrt{\left( \frac{1}{8}+\gamma \right)  \left(1-z_{{m}} \right)
\sqrt { \left( 1-z_{{m}} \right)  \left( 1+8\,\gamma \right)
 \left( -1+8\,\gamma \right) \Delta\, \left( -2\,z_{{m}}+\Delta\,z_{{m
}}-\Delta \right) }\pi }}.
\end{eqnarray}

We deduce  that $\langle{\cal O}_{+}\rangle$ vanishes at $T=T_c$,and
superconducting phase locates $T<T_c$. The mean field theory result
$\langle{\cal O}_{+}\rangle \propto(1-T/T_c)^{1/2}$
 revisited.
The value of  $T_c$  via (\ref{Tc}) is evaluated as $T_c\propto\sqrt[3]{\rho}$ is comparable with \cite{epl2}.

\section{Conclusions}
In the present paper, we completed our former studies about the
properties of Weyl's corrected p-wave holographic superconductors in
5-dimensional $AdS$-Schwarzschild background. Firstly, we briefly
clarified the key notes of gauge/gravity and how we relate the
solutions of gravity to a quantity in field theory. The equations of
motion in our interesting problem are nonlinear and coupled
differential equations, so one cannot obtain analytic solutions to
these equations in the closed form. Here we use the matching method.
In this method, we match the asymptotic AdS solutions to the horizon
solutions at a mid point, and then we find the expectation value of
the dual operator $\langle{\cal O}_{+}\rangle$ and the critical
temperature $T_c$ analytically. Surprisingly inspired from the string theory, the critical exponent of the condensation is independence from the Weyl correcxtions. These analytical results coincides completely on the previously numerical data \cite{epl2}.

\end{document}